\documentclass[preprint,preprintnumbers,amsmath,amssymb,nofootinbib]{revtex4}
\usepackage{graphicx,color}

\newcommand\nn{\nonumber \\}

\newcommand{\lk}{\left(}
\newcommand{\rk}{\right)}
\newcommand{\ltk}{\left\{}
\newcommand{\rtk}{\right\}}
\newcommand{\ldk}{\left[}

\newcommand{\rdk}{\right]}
\newcommand\beq{ \begin{eqnarray} }
\newcommand\eeq{ \end{eqnarray} }

\begin{document}


\title{Bose-Einstein condensation and density collapse 
in a weakly coupled boson-fermion mixture}
\author{Kyoko Shirasaki} 
\email{b12d6a05@s.kochi-u.ac.jp} 
\author{Eiji Nakano}
\email{e.nakano@kochi-u.ac.jp}
\affiliation{Department of Physics, Kochi University, 
Kochi 780-8520, Japan}
\author{Hiroyuki Yabu} 
\email{yabu@se.ritsumei.ac.jp}
\affiliation{Department of Physics,
Ritsumeikan University, Kusatsu 525-8577, Siga, Japan}

\date{\today}

\begin{abstract}
We investigate the mechanical instability toward density collapse 
and the transition temperature of Bose-Einstein condensation 
in the weakly coupled boson-fermion many-body mixture 
of single-component boson and fermion gases, 
in the case of the repulsive boson-boson and attractive boson-fermion interactions;
no fermion-fermion interaction is assumed due to the Pauli exclusion 
because of the diluteness of the mixture. 
The mechanical instability occurs in the part of bosonic gas 
in the case where the induced boson-boson attraction mediated by the fermion polarization overcomes 
the boson-boson repulsive interaction at low temperatures. 
We calculate the onset temperature of the mechanical instability 
and the BEC transition temperature 
for the interaction strength and the boson-fermion number ratio. 
We also discuss the relation between the mechanical instability 
and the BEC transition using a diagrammatic method. 
\end{abstract}


\maketitle
\section{Introduction}
Bose-Einstein condensation (BEC) provides a concept of fundamental importance in physics, 
and can be observed in a variety of phenomena, e.g., 
superfluidity of $^4$He, 
the spontaneous symmetry breaking due to Higgs boson condensation.  
In recent decades 
the development of optical techniques has enabled us to realize the BEC experimentally 
in a more accessible way \cite{Pitaevskii1,PethickSmith1}, 
where magnetically-trapped atomic gases are cooled down 
to form condensed states in the lowest energy level. 
Such systems are highly controllable in the sense that 
many species of atoms can be trapped together, 
and their interactions are tunable using Feshbach resonance. 
It is also possible to create the low-dimensional systems effectively, 
and to make optical lattice potentials 
by the interference of laser beams creating periodic potentials 
\cite{Pitaevskii1,PethickSmith1}. 

In the present paper we focus on the boson-fermion mixture 
of single-component boson and fermion gases,
where the boson-boson (boson-fermion) interacting is weakly repulsive (attractive). 
We assume that the gas of mixture is very dilute and cold, 
so that these interactions are characterized 
by the s-wave scattering lengths in the boson-boson and boson-fermion interactions, 
and no fermion-fermion interaction is assumed due to Pauli exclusion principle. 
Relevant systems can be observed in the trapped cold atom experiments 
\cite{Truscott1,Modugno}, 
and there are pioneering theoretical studies 
on ground state properties of the mixture in the trap: 
phase separations and density profiles \cite{Mlmer1,Miyakawa2}, 
and collective excitations and induced instabilities \cite{Miyakawa1.5, Miyakawa1,SMSY}.  

In this paper we study the many-body quantum theoretical aspects of the boson-fermion mixtures,  
especially the BEC transition temperature and the mechanical collapse of the boson sector 
under the influence of the fermion gas.
The theoretical study on the BEC transition temperature $T_c$ 
has a history for the uniform system of interacting Bose gas 
\cite{Pitaevskii1,Baym1}; 
the evaluation of the leading order shift of the $T_c$ 
due to the weak repulsion has long been a controversial subject. 
The state of the art calculations \cite{Baym1,Baym3} reveal 
that the leading order effect is linear in the boson-boson scattering length $a_{bb}$, 
\beq
 \Delta T_c\equiv \frac{T_c-T_0}{T_0}= c a_{bb}n_b^{1/3}, 
 \label{BECTc1}
\eeq
where $T_0$ and $a_{bb}$ are the free boson transition temperature and 
the boson-boson s-wave scattering length, respectively, and 
the positive numerical factor $c$ is of the order of unity. 
The above result shows that the critical temperature rises up 
with small and positive $a_{bb}$ in the leading order effect. 
Various non-perturbative approaches, 
which include renormalization group methods and numerical simulations, 
are used to evaluate the shift of critical temperature to the leading and sub-leading orders 
\cite{Stoof2,Arnold1}. 

In the trapped cold-gas experiments, 
the shifts of the transition temperature have been observed from the thermodynamical limit 
of the harmonic oscillator potential \cite{Pitaevskii1};
they are largely explained by the volume-expansion \cite{Giorgini4}
and the finite particle-number \cite{Grossmann1} effects, 
which do not exist in the uniform system 
but give larger effects in the trapped system.

The instability of the BEC of the attractive interaction had been studied in the uniform system \cite{Stoof1}.
Research interest has been generated by the experimental success of the attractive BEC of Li atoms \cite{Bradly} 
and the Bose Nova controlled by the Feshbach resonance method \cite{Donley}, 
where the quantum mechanical zero-point fluctuations support the BEC against the mechanical collapse 
by the attractive interaction \cite{BaymPethick,Houbiers,PethickSmith1}.
The many-body theoretical calculations have been done for the uniform attractive BEC \cite{Stoof1,Koh1,Houbiers,Mueller1}.

Now we turn to the low temperature behavior of boson-fermion mixtures, 
where the collapse of the trapped boson gas has been studied theoretically in the local-density 
and mean-field approximations \cite{Mlmer1,Miyakawa2,Miyakawa1.5,Miyakawa1,Tsurumi}.
The experimental performance of the trapped boson-fermion mixture was studied 
in the ${}^{40}$K-${}^{87}$Rb system \cite{Modugno}, 
and theoretical studies on the experiment has been done \cite{Capuzzi1,Roth1}.

In the uniform system of such boson-fermion mixture, 
one expects that the bosonic sector undergoes the BEC transition 
with temperature being lowered, 
although there must be a small shift of transition temperature 
from that of the free Bose gas due to the weak interactions. 
In the presence of fermions, however, 
the boson-boson effective interaction mediated by the fermion polarization 
becomes attractive regardless of the sign of the boson-fermion interaction 
\cite{Fetter1}. 
One also expects that there appears competition between the BEC 
and the collapse in the gas of mixture  
due to the influence of environmental fermions. 

For the trapped system  
theoretical studies on the BEC of the boson-fermion gaseous mixture 
have already been done \cite{Albus1,Hu1}, 
where the trap is assumed to be large enough so that finite size effects are ignorable, 
and the local density approximation is used to find 
that the shift of critical temperature comes mainly from density profiles in the trap. 
Thus, this shift is not a pure many-body effect. 

From the above experimental and theoretical background 
it is of our interest to obtain a precise many-body picture 
of the gaseous boson-fermion mixture in the weak coupling regime. 
In this paper 
we figure out how the BEC transition temperature 
(\ref{BECTc1}) is modified by the boson-fermion interaction, 
and where the density collapse induced by fermions takes place 
in the phase diagram of interaction strength and of thermodynamic variables. 


\section{Model Hamiltonian}
Let us consider the boson-fermion mixture in the uniform system. 
In terms of fermion (boson) annihilation operator $a_p$ ($b_p$) with momentum $p$, 
the effective Hamiltonian is given by, 
\beq
\mathcal{H}&=& \int_p \lk \xi_p a_p^\dagger a_p + \varepsilon_p b_p^\dagger b_p \rk 
+\int_k \int_p \int_q \lk g_{bf} a_{p+q}^\dagger b_{k-q}^\dagger b_k a_p
+g_{bb}  b_{p+q}^\dagger b^\dagger_{k-q} b_{k} b_{p}\rk, 
\label{Hamil1}
\eeq
where the boson and fermion single-particle energies are
$\varepsilon_p = \frac{p^2}{2m_b}-\mu_b$ and 
$\xi_p = \frac{p^2}{2m_f}-\mu_f$ 
with the free boson and fermion masses $m_b$ and $m_f$ 
and the chemical potentials $\mu_b$ and $\mu_f$, respectively. 
Note that the abbreviated notation is used for the momentum-space integration
$\int_p\equiv \int \frac{{\rm d}^3{p}}{(2\pi)^3}$ in the present paper.
The boson-boson and boson-fermion coupling constants are $g_{bb}$ and $g_{bf}$, respectively  
which are represented by the s-wave scattering lengths, $a_{bb}$ and $a_{bf}$. 
In the T-matrix approach, they become
\cite{PethickSmith1}, 
\beq
\frac{m_{ij}}{2\pi a_{ij}}
=
\frac{1}{g_{ij}}+\int_p \frac{1}{p^2/(2m_i)+p^2/(2m_j)}, 
\label{Tmatrix}
\eeq
where $\ltk i,j\rtk=\ltk b,f\rtk$, and 
$m_{ij}=\frac{m_im_j}{m_i+m_j}$ the reduced mass of particles $i$ and $j$. 
At the weak coupling regime, simply 
$g_{bb}=\frac{4\pi}{m_b}a_{bb}$ and $g_{bf}=\frac{2\pi}{m_{bf}}a_{bf}$. 
The above formulation is valid only for systems 
with a mean interparticle distance much larger 
than a typical size of particles $\sim r_0$, 
which sets the momentum cutoff in (\ref{Tmatrix}) to be $\sim 1/r_0$.

\section{BEC transition temperature}
In general, 
the BEC transition temperature $T_c$ of the uniform system is 
obtained by the criterion \cite{Thouless1}, 
\beq
\Sigma_0 -\mu_b=0
\eeq
where $\Sigma_0$ represents the boson self-energy at vanishing energy-momentum. 
In terms of the effective potential of the order parameter, 
$\langle \hat{\phi} \rangle$ obtained from the boson field operator $\hat{\phi}(x)$, 
the criterion implies that the second order coefficient vanishes and 
the system undergoes a continuous phase transition at the temperature. 
Note here that we have simply assumed that the present system 
exhibits the transition of the same order as in the weakly coupled Bose gas. 
As shown later, the effect of boson-fermion interaction mainly appears 
as the small modification of effective boson-boson interaction.

Since we are interested in the transition temperature 
at fixed densities of bosons and fermions, 
it is convenient to start with boson density $n_b$ 
at temperature $T$, 
\beq
n_b(T)=-T\int_k \sum_n G(\omega_n,k), 
\eeq
where 
the thermal Green's function $G(\omega_n,k)$ for the interacting boson is given by 
\beq
G^{-1}(\omega_n,k)=\frac{k^2}{2m_b}+\Sigma(\omega_n,k)-\mu_b 
\eeq 
with the self-energy $\Sigma(\omega_n,k)$, 
for the Matsubara frequency for bosons $\omega_n=2\pi n T$ and momentum $k$. 
From the argument in \cite{Baym1}, 
the $T_c$ is converted from the critical density, 
\beq 
n_b(T_c)-n_b^0(T_c)=
-T_c\int_k \sum_n \ldk G(\omega_n,k)-G^0(\omega_n,k) \rdk,
\label{eQy1}
\eeq
where $n_b^0(T)$ and $G^0(\omega_n,k)$ are the density 
and the Green function for the free boson, respectively. 
The free-boson density 
has the analytic expression at the critical point $\mu_b=0$: 
\beq
n_b^0(T)=\int_k (e^{k^2/2m_bT}-1)^{-1}=\zeta(3/2)/\lambda_{T}^3
\eeq
where 
$\zeta(3/2)=2.612$ 
and $\lambda_T=\sqrt{2\pi/mT}$ is the thermal wavelength. 
Using this relation, 
we obtain $n_b^0(T_c)=(T_c/T_0)^{3/2}n^0(T_0)$ 
where $T_0$ is the critical temperature of free bosons.  
Under the assumption that 
the critical densities of interacting and non-interacting bosons are equivalent: 
$n_b(T_c)=n_b^0(T_0)={\rm const.}$, 
we obtain
\beq
n_b(T_c)-n_b^0(T_c)&\simeq& 
 -\frac{3}{2}\frac{\Delta T_c}{T_0}n_b^0(T_0), 
\eeq
where $\Delta T_c=T_c-T_0$ is 
the shift of the transition temperature.
Combining this with (\ref{eQy1}), 
we obtain 
\beq
\frac{\Delta T_c}{T_0}
&=&
\frac{2T_0}{3n_b}\int_k \ldk G(0,k)-G^0(0,k) \rdk 
\nn
&=&
\frac{4m_bT_0}{3n_b}\int_0^\infty \frac{{\rm d}k}{2\pi^2}
\frac{2m_b\lk\Sigma_k-\Sigma_0\rk}{k^2+2m_b\lk\Sigma_k-\Sigma_0\rk},
\eeq
where we have taken the term of zero Matsubara frequency ($n=0$), and defined $\Sigma_k\equiv\Sigma(0,k)$. 
The zero Matsubara-frequency term gives a infrared (IR) singular contribution at the critical point, 
while the other terms  give the higher-order contributions for the coupling constant 
because the finite frequencies serve as the IR cutoff at zero momentum \cite{Baym1}. 
Thus we need to evaluate the boson self-energy, $\Sigma_k-\Sigma_0$, 
at the critical point in the boson-fermion mixture. 
\subsection{Boson self-energy to leading order}
Here we redefine the boson chemical potential 
as that measured from Hartree-Fock (mean-field) contributions, 
which have no momentum dependence:
\beq
\mu_b-\Sigma_{HF}  \rightarrow \mu_b. 
\eeq
It gives no influences on the value of the critical temperature within the mean-field approximation, 
since the critical point is still determined by 
$\mu_b-\Sigma_{HF}=0$. 

Considering the states very close to the critical point where $\mu_b=0$, 
we evaluate the dominant contributions of boson self-energy as in \cite{Baym1}.
\begin{figure}
  \begin{center}
   \includegraphics{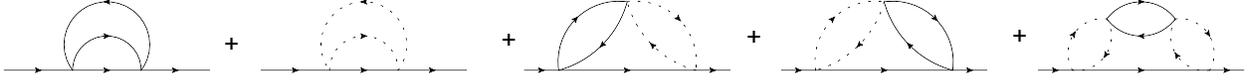}
    \caption{Leading diagrams contributing to the low momentum structure 
    of the boson self-energy at the critical point. 
    Solid (dashed) line implies the boson (fermion) propagator. }
  \label{fig1}
   \end{center}
\end{figure}
\subsubsection{Single boson bubble contribution} 
The leading contribution in the boson self-energy is obtained 
from the boson single-bubble diagram 
(the first diagram in Fig.~\ref{fig1}):  
\beq
\Sigma^{(1)} \lk \omega_l, k \rk 
&=& 
2g_{bb}^2 T\sum_n \int_q T\sum_m \int_p 
\frac{1}{\lk i\omega_m-\varepsilon_p \rk 
\lk i{\omega}_{m+n}-\varepsilon_{p+q}\rk }
\frac{1}{i\omega_{n+l}-\varepsilon_{q+k} }
\nn
&=&
2g_{bb}^2 \int_q  \int_p \frac{ n(\varepsilon_{q+k}) 
\lk n(\varepsilon_{p+q})-n(\varepsilon_p)\rk +n(\varepsilon_{p+q})\lk 1+n(\varepsilon_p) \rk}
{i\omega_l+\varepsilon_{p+q}-\varepsilon_p-\varepsilon_{q+k}  },
\eeq
where 
$n(x)=(e^{x/T}-1)^{-1}$ is the Bose distribution function. 
The IR singular contribution at the critical point reads, 
after the replacement $n(x) \rightarrow T/x$ 
which is valid around the critical point $\mu_b= 0$, 
\beq
\Sigma^{(1)} \lk 0, k  \rk 
&\simeq & 
-2g_{bb}^2 \int_q  \int_p \frac{T^2}
{\varepsilon_{p+q}\varepsilon_p\varepsilon_{q+k}}
=
-2g_{bb}^2 T\int_q B(q) 
\frac{1}{\varepsilon_{q+k}},
\label{singleboson1}
\eeq
where the boson bubble contribution $B(q)$ is given by
\beq
B(q)=\int_p \frac{T}{\varepsilon_{p+q}\varepsilon_p}.
\label{B1}
\eeq 
Because of the divergent behavior $B(q) \sim q^{-1}$ around $q =0$, 
which is clear from  the dimensional counting, 
the self-energy $\Sigma^{(1)} \lk 0, k \rk$ becomes logarithmically divergent at $k=0$. 

\subsubsection{Contribution from fermion polarization}
The lowest order contribution from the boson-fermion coupling comes from the fermion polarization
(the second diagram in Fig.~\ref{fig1}), 
\beq
\Sigma^{(2)} \lk \omega_l, k \rk 
&=&  g_{bf}^2 T\sum_n \int_q T\sum_m \int_p 
\frac{-1}{\lk i\bar{\omega}_m-\xi_p \rk \lk i\bar{\omega}_{m+n}-\xi_{p+q}\rk }
\frac{1}{i\omega_{n+l}-\varepsilon_{q+k} }
\nn
&=&  g_{bf}^2 \int_q  \int_p 
\frac{f(\xi_{p+q})-f(\xi_{p})}{i\omega_l+\xi_{p+q}-\xi_p -\varepsilon_{q+k} }
\lk n(\xi_{p+q}-\xi_{p})-n(\varepsilon_{q+k})\rk, 
\eeq
where $\bar{\omega}_m=(2m+1)\pi T$ is the Matsubara frequency for fermions, 
and $f(x)=(e^{x/T}+1)^{-1}$ is the Fermi distribution function.
Around the critical point the leading contribution comes from the term of the zero Matsubara frequency, 
and the low momentum part; using $n(x) \rightarrow T/x$ as in the boson bubble case,
we obtain
\beq
\Sigma^{(2)} \lk 0, k\rk
&\simeq& -g_{bf}^2 \int_q  \int_p \frac{f(\xi_{p+q})-f(\xi_{p})}{\xi_{p+q}-\xi_p}
\frac{T}{\varepsilon_{q+k}}
= Tg_{bf}^2 \int_q  \Pi\lk q\rk
\frac{1}{\varepsilon_{q+k}}. 
\eeq
The contribution of the fermion polarization, 
$\Pi\lk q\rk=\int_p \frac{f(\xi_{p+q})-f(\xi_{p})}{\xi_p-\xi_{p+q}}\sim q^0$ for small $q$,  
to the boson self-energy is IR regular, 
so that it plays a less important role than that from the boson bubble at the critical point
\footnote{See appendix~A for the detailed behavior of the polarization function. }.
\subsubsection{Boson bubble with insertion of fermion polarization}
\begin{figure}
  \begin{center}
      \resizebox{80mm}{!}{\includegraphics{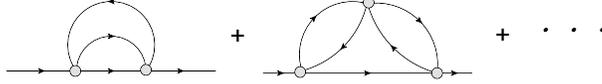}}
    \caption{Boson bubbles with an effective 
    boson-boson interaction (solid circle) including fermion polarization.}
    \label{fig2}
  \end{center}
\end{figure}

The above analysis shows that 
the single fermion polarization does not contribute 
to the leading IR divergence near the critical point. 
Thus, the boson bubbles with the fermion-polarization insertion 
have the same kind of the IR singularity as those without the insertion. 
Such contributions, which is obtained in the summation in the Fig.~\ref{fig1} 
except for the second diagram, 
are summarized as follows: 
\beq
\Sigma^{eff} \lk 0, k \rk 
&\simeq & 
-2T\int_q B(q) \ldk g_{bb}^2  -2g_{bb}g_{fb}^2 \Pi\lk q \rk 
+ \ltk g_{bf}^2 \Pi\lk q \rk \rtk^2 \rdk
\frac{1}{\varepsilon_{q+k}} 
\nn 
&=&
-2T\int_q B(q) \ldk g_{bb} -g_{bf}^2 \Pi\lk q \rk \rdk^2
\frac{1}{\varepsilon_{q+k}},
\label{selfeff}
\eeq
which has the same form with the single boson bubble 
with the effective boson-boson coupling:
\begin{equation}
g_{eff}(q)\equiv g_{bb} -g_{bf}^2 \Pi\lk q \rk. 
\label{bbeff1}
\end{equation}
It is represented as the first diagram in Fig.~\ref{fig2}. 
\subsection{Hardening of boson energy spectrum and $\Delta T_c$}
In the na{\"i}ve perturbation calculation, 
the boson self-energy is found to have the IR singularity, 
which should be attributed to the onset of a critical region in low momentum space;
the Ginzburg criterion tells us 
that the perturbative treatments break down 
for the interactions among long wavelength fluctuations \cite{Ginzburg1}. 
In order to remedy such a singularity, 
one needs to employ at least non-perturbative 
treatments even if the coupling constants are perturbatively small. 
The self-consistent method is one of such non-perturbative treatments, 
where a modification of low energy spectrum in the critical region 
is treated self-consistently \cite{Landau1,Patashinskii1,Baym1}. 
\footnote{
Another possibility is a resummation method, 
which will be used in the discussion later on.}
In this paper  
we evaluate the low momentum structure of boson single-particle energy, 
and see that the spectrum for lower momenta becomes hard 
in the self-consistent treatment. 

The dominant contribution of the self-energy 
$\Sigma^{eff}(0, k)\equiv \Sigma_k^{eff}$ for the small momentum $k$ 
comes from the small momentum region in the loop integral:
\beq
2m_b\lk\Sigma_k^{eff}-\Sigma_0^{eff}\rk
&=&
-4m_bT\int_q B(q) \ldk g_{bb}  -g_{bf}^2 \Pi\lk q \rk \rdk^2
\ldk \frac{1}{\varepsilon_{q+k}}-\frac{1}{\varepsilon_{q}} \rdk 
\nn 
&\simeq&
-4m_bT\int_q  \ldk g_{bb} -g_{bf}^2 \Pi\lk 0 \rk \rdk^2
\frac{ B(q-k) -B(q)}{\varepsilon_{q}}. 
\label{effself1}
\eeq
In the region where the momentum is smaller than a critical scale $k_c$, 
the boson single-particle energy is assumed to be modified as $\sim k^{2-\eta}$, 
where the exponent $\eta$ is an anomalous dimension;
the modification comes from 
the behavior of the order-parameter correlation-function in momentum space. 
On the basis of the argument in \cite{Patashinskii1,Baym1}, 
we replace the mean-field boson single-particle energy, 
$\varepsilon_k=k^2/2m_b-\mu_b$, 
appearing in Eqs.~(\ref{B1}) and (\ref{effself1}), 
with a self-consistent energy: 
\beq
\varepsilon_k=\frac{k^2}{2m_b}+\Sigma_k^{eff}-\Sigma_0^{eff}
=\frac{k_c^{1/2}k^{3/2}}{2m_b}\theta(k_c-k)
+\frac{k^2}{2m_b}\theta(k-k_c),
\label{effenergy1}
\eeq
and $\Sigma_0^{eff}-\mu_b=0$ at the critical point.
Here the exponent $3/2$ comes from a simple power counting estimation for loop integrals: 
Assuming a low-momentum behavior of the single-particle energy as $\varepsilon_k \sim k^\alpha$ 
in Eq.~(\ref{effself1}), 
one obtains the self-consistent equation 
$k^{6-3\alpha}=k^\alpha$,
which is satisfied with $\alpha=3/2$ \cite{Patashinskii1}. 

Actually we can check the self-consistency of Eq.~(\ref{effenergy1}) in the region of $k \simeq 0$. 
Substituting it explicitly into Eq.~(\ref{effself1}) and taking the leading order term, 
we obtain\footnote{For the details, see appendix~B.}
\beq
\Sigma_k^{eff}-\Sigma_0^{eff}
\simeq 
\ldk g_{bb}  -g_{bf}^2 \Pi\lk 0\rk \rdk^2
\frac{16m_b^3T^2}{15\pi^3}\lk \frac{k}{k_c}\rk^{3/2}. 
\label{effself2}
\eeq
In comparison of it with Eq.~(\ref{effenergy1}), 
the onset momentum scale $k_c$ is found to be 
\beq
k_c &=& 
\left[ g_{bb}  -g_{bf}^2 \Pi\lk 0\rk \right]
\sqrt{\frac{32m_b^4T^2}{15\pi^3}} 
\nonumber\\
&=& 
\left[ \frac{4\pi a_{bb}}{m_b} 
-\lk \frac{2\pi a_{bf}}{m_{bf}}\rk^2 \Pi\lk 0\rk \right]
\sqrt{\frac{2}{15\pi}}\frac{4 m_b^2T}{\pi}. 
\label{kcT}
\eeq

Now we estimate the transition temperature $T_c$ of the interacting mixture 
with the modified dispersion relation (\ref{effenergy1}). 
Substituting it into the free-boson distribution function at the critical point:
\beq
n_b(T_c)
=\int_p \frac{1}{e^{\varepsilon_p/T_c}-1}
\simeq 
n_b^0(T_c)-\frac{m_bT_c}{3\pi^2}k_c, 
\eeq
we obtain the shift of $T_c$:
\beq
\frac{\Delta T_c}{T_0}
\simeq
\frac{2m_bT_0}{9n_b\pi^2}k_c, 
\label{Tc1}
\eeq
where $k_c$ is given by Eq.~(\ref{kcT}) for $T=T_0$. 

As the other method to introduce the $k^{3/2}$ behavior of the single-particle energy,
we take the ansatz for the self-energy shown in \cite{Baym1},  
\beq
2m_b\lk\Sigma_k^{eff}-\Sigma_0^{eff}\rk
=\frac{k_c^{1/2}k^{3/2}}{1+\lk k/k_c\rk^{3/2}}, 
\eeq
which smoothly connects the low and high momentum regions. 
This ansatz leads to the critical temperature shift as 
\beq
\frac{\Delta T_c}{T_0}
&=&
\frac{4m_bT_0}{3n_b}\int_0^\infty \frac{{\rm d}k}{2\pi^2}
\frac{2m_b\lk\Sigma_k-\Sigma_0\rk}{k^2+2m_b\lk\Sigma_k-\Sigma_0\rk}
\simeq
\frac{2m_bT_0}{3n_b\pi^2} 1.184 k_c.
\label{Tc2}
\eeq
The difference between (\ref{Tc1}) and (\ref{Tc2}) is only on a numerical factor, 
which is of the order of unity. 

The hardening of the energy spectrum is understood as follows: 
at low temperatures 
the low-energy bosons, which have large quantum fluctuations because of large wavelengths, 
begin to overlap but repels each other by the repulsive interactions between them; 
it results in the decrease of the density of states in the low-energy region. 
It makes the accommodation of particles of the low-energy states easier at the high temperature, 
and explains the positive shift of $T_c$. 

Note here that  
the finite and trapped systems have 
the mechanism of the transition temperature shift, 
which is different from the many-body effect 
as discussed in \cite{Grossmann1,Giorgini4}.

\section{Density correlation function and mechanical instability}
We turn to the investigation of the mechanical instability toward the density collapse of the system; 
the instability starts first in the long-wavelength density fluctuations. 
In order to obtain the critical condition, 
we study the second order coefficients of the thermodynamic potential $F$ 
in terms of the boson and fermion densities:
\beq
M\equiv 
\frac{1}{2}
\lk 
\begin{array}{cc}
\frac{\partial^2 F}{\partial n_b^2 } & \frac{\partial^2 F}{\partial n_b\partial n_f}\\
\frac{\partial^2 F}{\partial n_f\partial n_b} & \frac{\partial^2 F}{\partial n_f^2}\\
\end{array}
\rk
=
\frac{1}{2}
\lk 
\begin{array}{cc}
\frac{\partial \mu_b}{\partial n_b} & \frac{\partial \mu_b}{\partial n_f}\\
\frac{\partial \mu_f}{\partial n_b} & \frac{\partial \mu_f}{\partial n_f}\\
\end{array}
\rk.
\eeq
Each component of the matrix $M$ is inversely proportional 
to the density-density correlation (response) function: 
\beq
\frac{\partial n_i}{\partial \mu_j}
=\langle n_i n_j\rangle, \quad \mbox{for}\quad i,j=b \hbox{ or } f, 
\eeq
which is related to the compressibility: 
$\kappa^{-1} =\left. -V\frac{\partial P}{\partial V}\right|_T 
=\sum_{i,j} n_in_j\frac{\partial \mu_i}{\partial n_j}$. 
The stability condition is equivalent 
to the positive definiteness of the eigenvalues: 
$M_{bb}>0$ (or $M_{ff}>0$) and ${\rm Det} M>0$. 
Thus, the instability occurs at either $M_{bb}=0$ or ${\rm Det} M=0$, 
where the compressibility diverges. 
In what follows we employ the random phase approximation (RPA) 
to evaluate the density-density correlation functions. 
\subsubsection{Boson-boson and boson-fermion type correlations}
We consider the Schwinger-Dyson-type equations 
for $\langle n_b n_b\rangle$ and $\langle n_b n_f\rangle$.
For the interaction kernel of the boson two-particle irreducible part, 
we should use  
\beq
\Gamma_{bb} \lk \omega_l, k\rk 
&=&  g_{bb} -g_{bf}^2\Pi\lk  \omega_l, k\rk, 
\eeq
where $\Pi\lk  \omega_l, k\rk$ is the fermion polarization function 
with finite energy-momentum\footnote{For details, see appendix~A}. 
The bare part for $\langle n_b n_b\rangle$ should be the single boson bubble, 
and that for $\langle n_b n_f\rangle$ 
be the single boson bubble multiplied by the single fermion polarization with the coupling constant $g_{bf}$. 
Then, the RPA type solutions are given by 
\beq
\langle n_b n_b\rangle 
&=& 
\frac{B(0,0)}{1+\Gamma_{bb}(0,0)B(0,0)} 
\label{bbcol}
\\
\mbox{and}\quad
\langle n_b n_f\rangle 
&=&
g_{bf}\Pi(0,0) \langle n_b n_b\rangle, 
\label{bfcol}
\eeq
where 
$B(0, 0)=\frac{m_b}{2\pi^2} \int_0^\infty {\rm d}p \, n(\varepsilon_p)$ 
is the low energy-momentum limit of 
single boson bubble function: 
\beq
B\lk \omega_n, q \rk 
&=& T\sum_m \int_p 
\frac{1}{\lk i\omega_m-\varepsilon_p \rk 
\lk i{\omega}_{m+n}-\varepsilon_{p+q}\rk }
\nn
&=& 
\int_p \frac{n(\varepsilon_{p+q})-n(\varepsilon_{p})}
{i\omega_n-\varepsilon_{p+q}+\varepsilon_p }, 
\eeq
and we have sent the external energy-momentum to zero, 
since $\langle n_i n_j \rangle$ corresponds to 
the correlation function in the long wavelength limit.  
\subsubsection{Fermion-fermion type correlation }
For the fermion density-density correlation $\langle n_fn_f\rangle$, 
the bare part is given simply by the fermion polarization, 
and the interaction kernel for the fermion two-particle irreducible part, $\Gamma_{ff}$, 
includes the infinite sum of boson bubbles, 
\beq
\Gamma_{ff}\lk \omega_l, k\rk 
&=&
-g_{bf}^2\frac{ B\lk \omega_l, k\rk }{1+g_{bb}B\lk \omega_l, k\rk }. 
\eeq
Then, the RPA type solution becomes 
\beq
\langle n_fn_f\rangle = \frac{\Pi(0,0)}{1+\Gamma_{ff}(0,0)\Pi(0,0)}
=\frac{\Pi(0,0)\lk 1+g_{bb}B(0,0)\rk }{1+\ldk g_{bb}-g_{bf}^2\Pi(0,0)\rdk B(0,0)}. 
\label{ffcol}
\eeq

In Eqs. (\ref{bbcol})-(\ref{bfcol}) and (\ref{ffcol}), 
the density-density correlation functions have a common denominator, 
so that the mechanical instability for long-range density fluctuation occurs at 
\beq
1+g_{eff}(0) B(0,0)=0, 
\label{RPA1}
\eeq
where $g_{eff}(0)$ is 
the effective static potential between bosons given in Eq.~(\ref{bbeff1}). 
\section{Result and discussion}
\begin{figure}
  \begin{center}
       \resizebox{50mm}{!}{\includegraphics{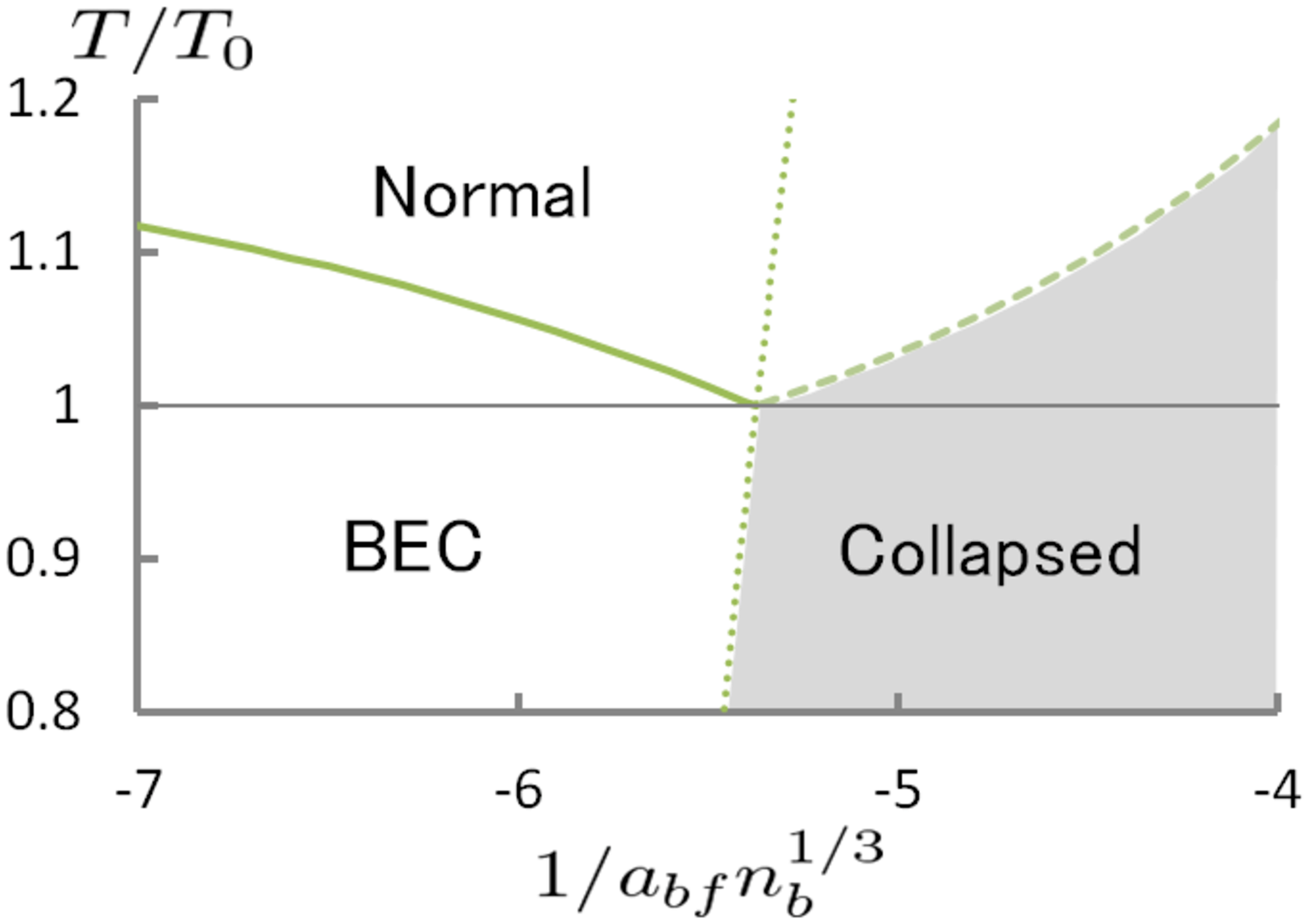}}
      \resizebox{50mm}{!}{\includegraphics{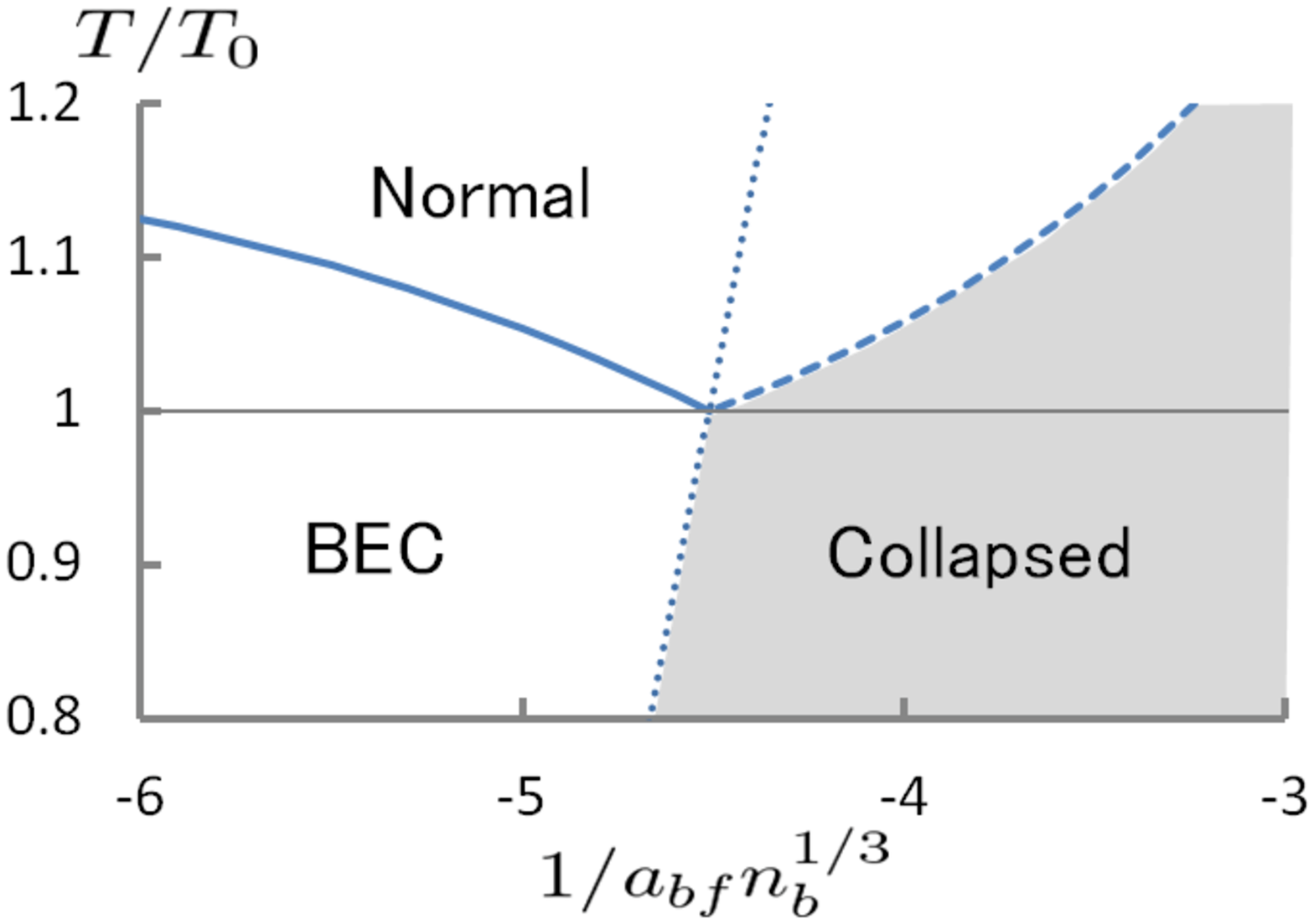}} 
      \resizebox{50mm}{!}{\includegraphics{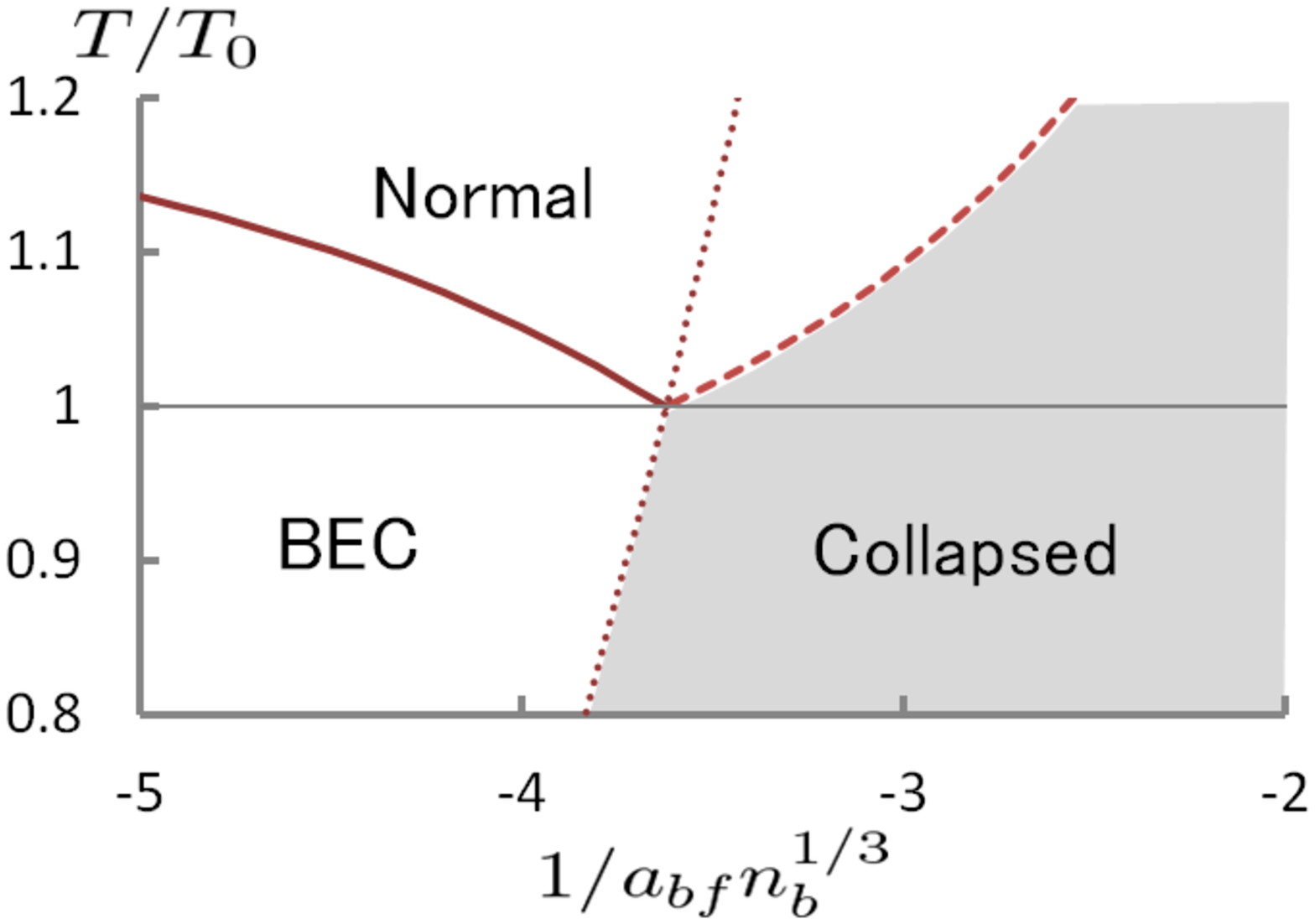}} \\
       \resizebox{50mm}{!}{\includegraphics{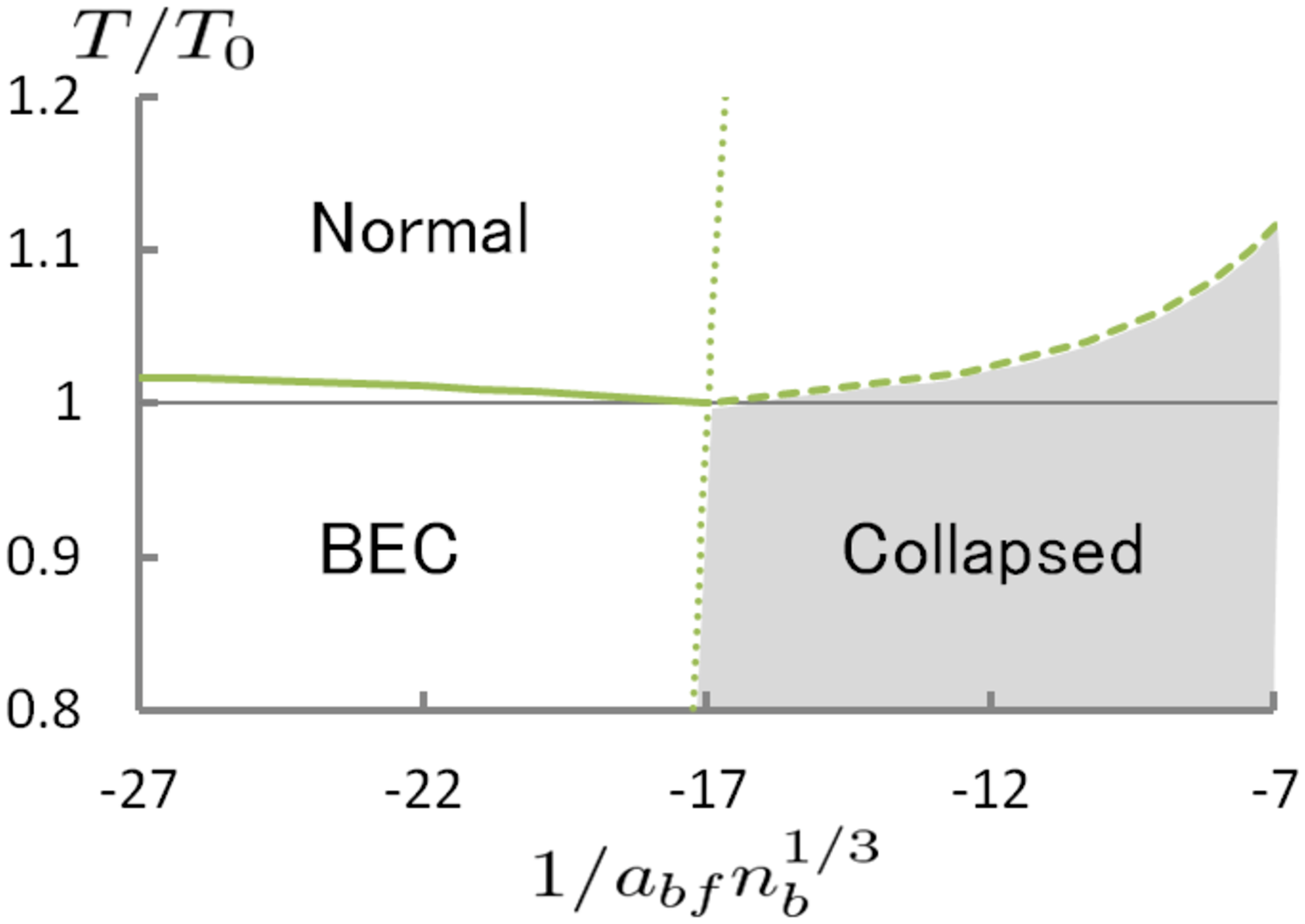}}
      \resizebox{50mm}{!}{\includegraphics{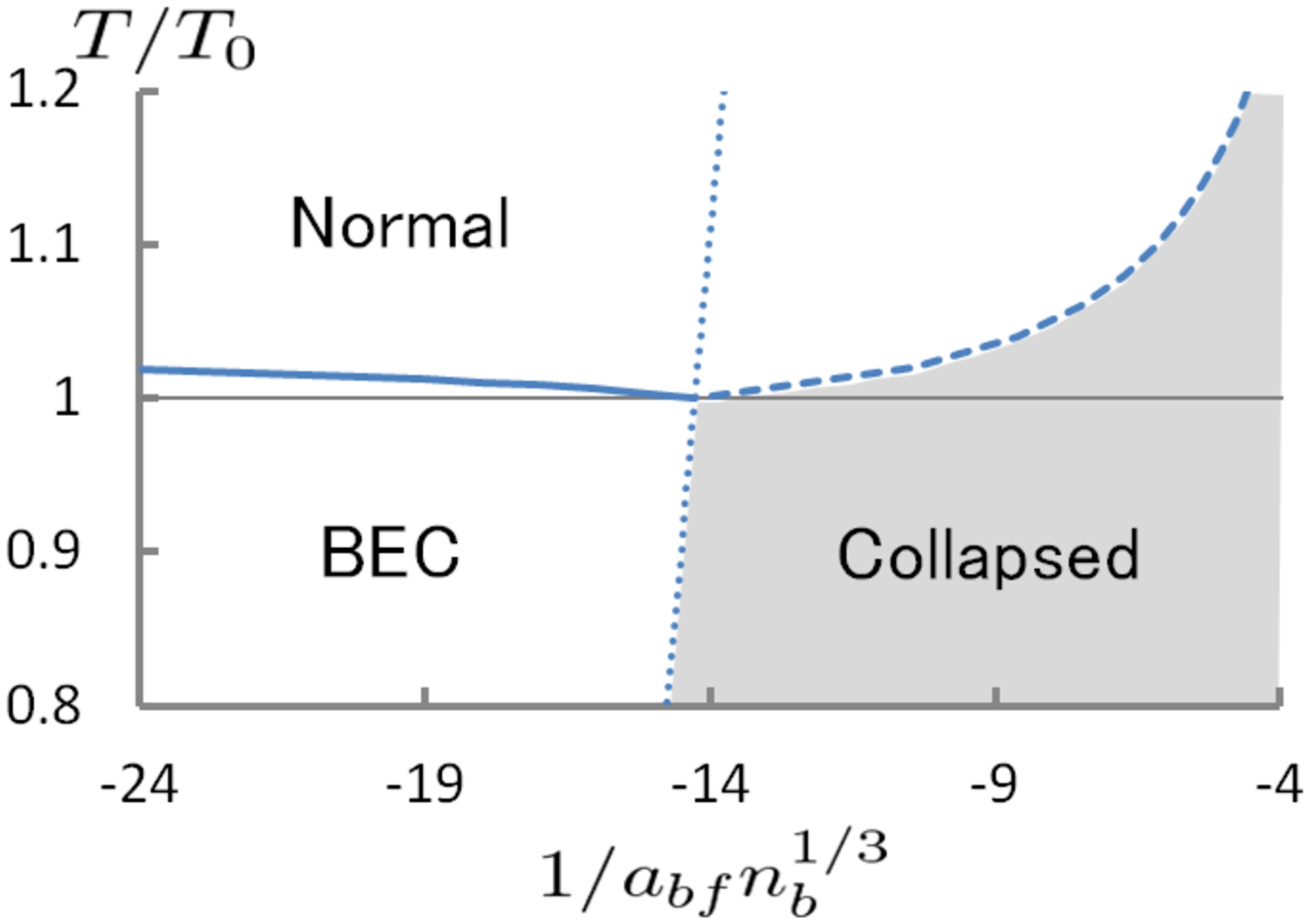}} 
      \resizebox{50mm}{!}{\includegraphics{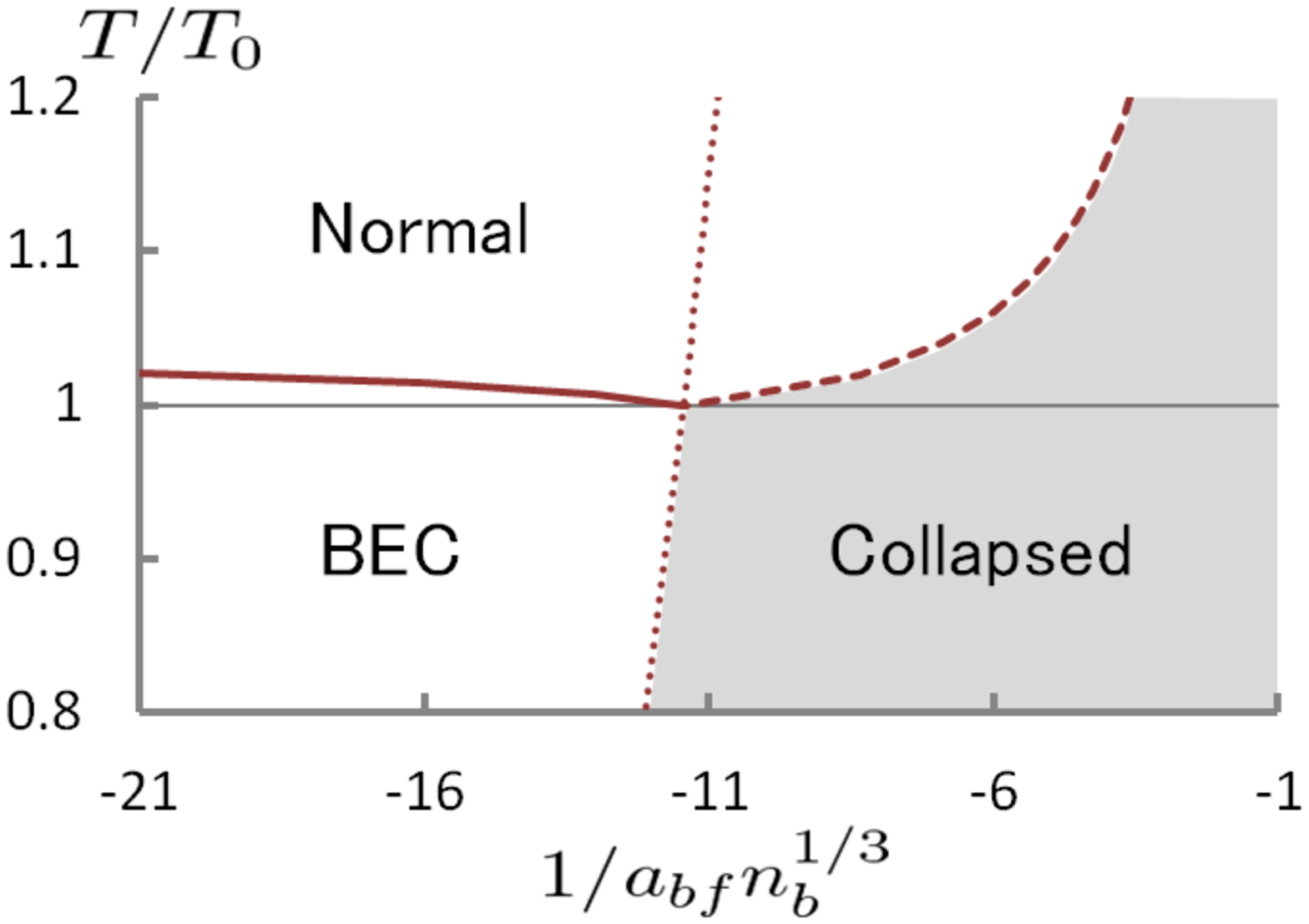}} \\
   \caption{(Color online) The phase diagrams of the boson-fermion mixture at a fixed boson density $n_b$: 
   the normalized BEC transition temperature (solid line), 
   the instability boundary to density collapse (dashed line), 
   and zeros of the effective boson-boson interaction 
   (dotted line), 
   for the inverse of dimensionless boson-fermion scattering length. 
   Here $T_0$ is the free boson BEC transition temperature, 
   $1/a_{bb}n_b^{1/3}=10$ (top row), $100$ (bottom row), and $m_f/m_b=1$. 
   From left to right $n_f/n_b=2,1,0.5$.
   }
  \label{fig3}
   \end{center}
\end{figure}

In Fig.~\ref{fig3}, 
we show the onset temperature of the mechanical instability (dashed line) 
as a function of the boson-fermion scattering length $a_{bf}$ normalized with a fixed boson density $n_b$ 
in the weak coupling region, 
for the fermion-boson density ratios $n_f/n_b =2$ (left), $1$ (center), and $0.5$ (right);
the onset temperature is obtained from the numerical solutions 
of Eq.~(\ref{RPA1}) \footnote{For the details, see appendix~C}. 
We also plot the BEC transition temperature $T_c$ (solid line) obtained from Eq.~(\ref{Tc2}). 
The dotted line is the temperature where the low energy effective coupling constant (\ref{bbeff1}) vanishes,
$g_{eff}(0)=g_{bb}-g_{bf}^2\Pi(0,0)=0$;
the $g_{eff}$ becomes positive(negative) above(below) the line.
These three lines meet at $T=T_0$: the BEC transition temperature of the corresponding free Bose gas. 
We find that the diagrams in Fig.~\ref{fig3} for the different density ratios have
no qualitative changes in the phase structure; 
this is because,
though the increase of $n_f/n_b$ enhances the fermion polarization,
the large part of its effect is to shift the intersection point on the line $T=T_0$, 
e.g., 
to $1/a_{bf}n_b^{1/3}=-5.37, -4.51, -3.62$ 
for $n_f/n_b=2,1,0.5$, respectively for $1/a_{bb}n_b^{1/3}=10$. 
We show results only for the values of boson-boson scattering length $1/a_{bb}n_b^{1/3}=10, 100$, 
since different values of it do not bring qualitative changes in the result. 
The intersection point of $1/a_{bf}n_b^{1/3}$ is only inversely proportional to $\sqrt{a_{bb}n_b^{1/3}}$, 
and the $T_c$ is suppressed by a factor $a_{bb}n_b^{1/3}$ for its small values. 
The calculations are for the mixture of the mass ratio $m_f/m_b=1$;
as shown in Appendix~C,
the dependence of the mass ratio should be absorbed 
into the boson-fermion scattering length and the fermion density. 

The BEC transition temperature obtained from Eq. (\ref{Tc2}) 
is not extended into the region of negative values of the effective coupling constant $g_{eff}(0)<0$, 
but a simple application of Eq. (\ref{Tc2}) 
gives the BEC transition temperature below the onset temperature of the mechanical instability. 
We can understand it as follows. 
As already mentioned on the BEC transition earlier, 
the IR singularity to the leading order of coupling constants 
comes from the single boson bubble contribution at the critical point, i.e., 
$B(q)=\int_p \frac{T}{\varepsilon_{p+q}\varepsilon_p}\sim 1/q$ 
in Eq.~(\ref{singleboson1}).  
The infinite resummation of the boson bubbles   
serves as an alternative 
to the self-consistent treatment to remedy the singularity \cite{Baym1}.
Actually the boson-bubble resummation 
with a positive effective coupling constant $g_{eff}(0)>0$, 
as in Fig.~\ref{fig2}, 
gives a screening effect for the single boson-bubble contribution,  
\beq
B(q) \rightarrow \frac{B(q)}{1+g_{eff}(q)B(q)}, 
\eeq
which is the IR-singularity-free at $q=0$ for the critical point of $\mu_b=0$. 
However, in the region of $g_{eff}(0)<0$, 
it works as the anti-screening effect, 
and a singular behavior occurs at $q=0$ on a certain temperature above the BEC transition 
where $\mu_b\neq 0$. 
This singular behavior is nothing but the onset of the mechanical instability
in the RPA treatment. 

The temperature of the mechanical instability
sets in at the intersection point $T=T_0$, 
\footnote{
The compressibility of the free boson gas, 
which corresponds to the single boson bubble, 
becomes divergent at the BEC transition temperature $T_0$.}
and rises up with the increasing $a_{bf}$,
because the larger thermal pressure is necessary 
for the Bose gas to overcome the mechanical instability 
coming from the strong attraction. 
In the temperature region below $T_0$, on the other hand, 
the instability points should be on the top of the curve of vanishing effective potential, 
since the condensate of bosons gives a divergent contribution 
in the density-density correlation at the vanishing external energy-momentum 
as shown in \cite{Griffin1,Mueller1}. 
The different situation exists in the case of the trapped pure Bose gases of size $L$ 
with the attractive boson-boson interaction $g_{bb}<0$; 
in this case, the zero-point quantum fluctuation energy of the order of $L^{-2}$ exists 
and it may overcome the mechanical instability 
when the boson local density is small enough,
i.e., 
$|g_{bb}|n_b < \pi^2/m_b L^2$ \cite{Mueller1}. 

In the trapped systems of the boson-fermion mixture, 
the symptom of the mechanical instability can be found 
in the collective excitation,  
where the energy of the breathing mode becomes soft 
when the boson-fermion attractive interaction approaches the critical value \cite{SMSY}.  

Note here that the Evans-Rashid transition \cite{Evans1}, 
that is, the BCS-type boson-boson pairing transition,
has been discussed as a possibility for the instability 
in the uniform and trapped systems for the boson gas of the attractive interaction \cite{Stoof1,Mueller1};
the order parameter of the transition is   
$\langle \hat{\phi} \hat{\phi} \rangle$.
In the Evans-Rashid transition, 
it is concluded that the condensation of the boson pairs  
takes place in a metastable region under the spinodal line. 
In the phase diagrams in  Fig.~\ref{fig3}, 
the corresponding Evans-Rashid transition by the effective attraction should 
appear in the small window between the mechanical instability temperature and $T=T_0$. 

We note also that a boson-fermion mixture similar to ours but with up-down fermions 
was studied for simulating dense QCD matter of strong coupling \cite{Maeda1}, 
where the same qualitative result for the $T_c$ of BEC was mentioned 
and a schematic phase diagram is presented 
for BSC-type fermion pairings as well as the BEC and the density collapse obtained from a semi-classical approximation. 
A quantum Monte Carlo simulation in three dimensions for a weakly coupled mixture with up-down fermions was also performed 
in \cite{Yamamoto1} to find that for fixed chemical potentials
the BEC (superfluid bosons) is enhanced through the effective chemical potentials with increasing boson-fermion coupling. 
Our present result gives a more precise many-body description and reason for the phase structure for the weak coupling regime 
in the absence of the fermion pairings and at given densities of bosons and fermions.

\section{Summary and outlook}
We have investigated the BEC transition and the mechanical instability
in the weakly coupled boson-fermion mixture of single-component boson and fermion gases 
with the repulsive boson-boson and the attractive boson-fermion interactions;
no interaction is assumed among fermions. 
We have evaluated the BEC transition temperature (\ref{Tc2}) 
using the self-consistent method by \cite{Baym1}, 
and the instability temperature 
calculated in the random phase approximation (\ref{RPA1}); 
the results are summarized in the phase diagrams in Fig.~\ref{fig3}.

As a characteristic feature in the leading order result 
the BEC transition and the instability meet at $T=T_0$ the free boson transition point, 
where the effective boson-boson coupling $g_{eff}(0)$ in Eq.~(\ref{bbeff1}) vanishes. 
At temperatures below $T_0$, the regions of BEC and mechanical instability separates 
by the boundary obtained as the vanishing effective boson-boson coupling constant. 
The region of the density collapse is 
beyond the description of our effective model (\ref{Hamil1}) for the dilute gas. 
 
The results obtained here are valid only in the weak coupling region 
of the boson-boson and boson-fermion interactions. 
In the strong coupling region, 
including the unitarity limit of boson-fermion scattering, 
some studies exist on the BEC transition and the instability,
mainly incorporating the boson-fermion pairing effects:
for example, the stability condition \cite{Yu1}, 
the boson-fermion paring and phase diagram \cite{Maeda1,Albus2}, 
and the suppression of BEC due to the pairing effects \cite{Powell1}. 
In such systems many-body correlations other than the boson-fermion pairing 
are also important to determine the phase structure, 
and, the inclusion of these effects is beyond the description of present approximations 
and should be an interesting study in future. 

KS and EN acknowledge useful comments 
from Shun-ichiro Koh, Yasuhiko Tsue, and Kei Iida.   


\appendix
\section{Fermion polarization function}
The fermion polarization function is given by 
\beq
\Pi\lk \bar{\omega}_n, q \rk 
&=& T\sum_m \int_p \frac{-1}{\lk i\bar{\omega}_m-\xi_p \rk 
\lk i\bar{\omega}_{m+n}-\xi_{p+q}\rk }
\nn
&=& \int_p \frac{f(\xi_{p+q})-f(\xi_{p})}{i\bar{\omega}_n-\xi_{p+q}+\xi_p }. 
\eeq
At $\bar{\omega}_n =0$, 
it becomes
\beq
\Pi\lk 0, q \rk 
&=&
\frac{2m_f}{q}\int_0^\Lambda \frac{{\rm d}p p}{(2 \pi)^2}f(\xi_{p}) 
\ln{\left|\frac{2p+q}{-2p+q}\right|}
\nn
&=&
\frac{m_f}{2 \pi^2}\int_0^\infty {\rm d}p 
F'_p  \frac{1}{q}\ln{\frac{|2p+q|}{|2p-q|}} 
= 
-\frac{m_f}{2 \pi^2}\int_{-\infty}^\infty {\rm d}p \frac{F_{p-q/2}}{q}\frac{1}{p}
\nn 
&=& 
\frac{m_f}{2 \pi^2}\int_{-\infty}^\infty {\rm d}p \frac{f(\xi_{p})}{2}
\Big[ 1 - \ldk 1-f(\xi_{p})\rdk \ltk 1+\frac{p^2}{3m_f T}(2f(\xi_{p})-1) \rtk \frac{q^2}{8 m_f T}
\nn
& &\qquad\qquad\qquad\qquad\qquad +\mathcal{O}\lk q^4\rk \Big], 
\eeq
where $F_p=m_f\ldk \frac{p^2}{2m_f}-\mu_f +T\ln f(\xi_p)\rdk$.

\section{Boson bubble and self-consistent single-particle energy }
The low momentum structure of the boson-bubble function 
at the critical point using the modified dispersion (\ref{effenergy1}) becomes
\beq
B(q)
&=&\int_p \frac{T}{\varepsilon_{p+q}\varepsilon_{p}}
=T \frac{2\pi (2m_b)^2}{k_c(2\pi)^3}\int_{0}^{k_c}{\rm d}p p^2
\int_{-1}^{1}{\rm d}x \frac{1}{(p^2+2qpx+q^2)^{3/4}p^{3/2}}
\nn
&=&T\frac{2m_b^2}{k_c\pi^2}\int_{0}^{k_c}{\rm d}p\, p^{1/2}
\frac{\sqrt{p+q}-\sqrt{|p-q|}}{pq}
\nn
&=&T\frac{2m_b^2}{k_c\pi^2}
\lk 1-\frac{\pi}{2}+2\ln 2+\ln{\frac{k_c}{q}}\rk +\mathcal{O}\lk q^2 \rk. 
\eeq

Substituting the above result into the boson self-energy,
we obtain
\beq
\Sigma_k-\Sigma_0
&=&
-2g_{eff}^2(0) T\int_q B(q) 
\ldk \frac{1}{\varepsilon_{k+q}}-\frac{1}{\varepsilon_{q}}\rdk
\nn
&=&
g_{eff}^2(0) \frac{16m_b^3T^2}{15\pi^3}\lk \frac{k}{k_c}\rk^{3/2}
+\mathcal{O}\lk k^2 \rk. 
\eeq
This shows the self-consistency of the low-momentum structure of the dispersion relation:
$\varepsilon_k = \frac{k^2}{2m_b}+\Sigma_k-\mu_b =  
\frac{p^2}{2m_b}+\Sigma_k-\Sigma_0 \sim k^{3/2}$ for $k\sim 0$,
where we used the integral formula: 
\beq
&&\int_q \ln\lk \frac{q}{k_c}\rk \left\{ \frac{1}{|q-k|^{3/2}}-\frac{1}{q^{3/2}} \right\} 
=
\int_q  q^{-3/2} \ln\frac{|q+k|}{|q|}
\nn
&=&
\int_0^{k_c} {\rm d}q\, q^{1/2} \int_{-1}^{1} {\rm d}x \frac{2\pi}{2(2\pi)^3}
\ln\frac{q^2+2kqx+k^2}{q^2}
\nn 
&=&
\frac{1}{2(2\pi)^2}\int_0^{k_c} {\rm d}q\, q^{1/2} \left\{
-2-\frac{(q-k)^2}{2kq}\ln\frac{(q-k)^2}{q^2}
+\frac{(q+k)^2}{2kq}\ln\frac{(q+k)^2}{q^2}\right\} 
\nn 
&=&
\frac{2}{15\pi}k^{3/2}+\mathcal{O}\lk k^2 \rk. 
\eeq

Note that the above calculation is a rederivation of the result in \cite{Baym1} 
except that the boson-boson coupling constant is the effective one, $g_{eff}(0)$, in the present case. 

\section{Normalized form of equations}
In order to draw the phase diagram 
for the temperature $T$ and boson-fermion scattering length $a_{bf}$
at fixed boson density $n_b$ and boson-boson scattering length $a_{bb}$,  
we represent the BEC transition temperature $T_c$ 
and the equation for the mechanical instability
with the dimensionless quantities: 
the fermion-boson density ratio $n_f/n_b$, 
$\eta_{bb}^{-1}\equiv a_{bb}n_b^{1/3}$, 
$\eta_{bf}^{-1}\equiv a_{bf}n_b^{1/3}$, and the temperature $T/T_0$, 
where $T_0=\frac{2\pi}{m_b}\left\{ \frac{n_b}{\zeta(3/2)} \right\}^{2/3}$.

The BEC transition temperature (\ref{Tc2}) scaled with $T_0$ is represented as 
\beq
\frac{\Delta T_c}{T_0}
&=&
\frac{128 c}{3 \sqrt{\pi} \left\{ \zeta(\frac{3}{2}) \right\}^{\frac{4}{3}}}
\sqrt{\frac{2}{15}}\left. \tilde{g}_{eff}\right|_{T=T_0},
\eeq
with a numerical constant  $c=1.18$,
and the scaled effective coupling constant $\tilde{g}_{eff}$:
\beq
\tilde{g}_{eff}
&=&
\eta_{bb}^{-1}
\ldk 1-\frac{\eta_{bb}}{\eta_{bf}^2}
\frac{1}{\sqrt{\pi}\ltk \zeta(\frac{3}{2})\rtk^{\frac{1}{3}}}
\lk \frac{m_f}{m_b}\rk^{3/2}\lk 1+\frac{m_b}{m_f}\rk^2 
\tilde{\Pi}_0\rdk
\\ 
\tilde{\Pi}_0
&=& \int_0^\infty {\rm d}x \lk e^{\frac{x^2-\mu_f/T_0}{T/T_0}}+1\rk^{-1}. 
\eeq

The deterministic equation (\ref{RPA1}) of the mechanical instability in RPA becomes 
\beq
0
&=&
1+\frac{4}{\sqrt{\pi}\ltk \zeta(\frac{3}{2})\rtk^{\frac{1}{3}}}
\tilde{g}_{eff} \tilde{B}_0, 
\\ 
\tilde{B}_0
&=& 
\int_0^\infty {\rm d}x \lk e^{\frac{x^2-\mu_b/T_0}{T/T_0}}-1\rk^{-1}.  
\eeq
The above equations are solved with the equations 
that relate the chemical potentials $\mu_{b,f}$ 
to the fermion-boson density ratio $n_f/n_b$, 
\beq
1
&=&
\frac{4}{\sqrt{\pi} \zeta\lk \frac{3}{2}\rk}
\int_0^\infty {\rm d}x\, x^2 \lk e^{\frac{x^2-\mu_b/T_0}{T/T_0}}-1\rk^{-1}, 
\\
\frac{n_f}{n_b}
&=& 
\lk \frac{m_f}{m_b}\rk^{3/2}
\frac{4}{\sqrt{\pi} \zeta\lk \frac{3}{2}\rk}
\int_0^\infty {\rm d}x\, x^2 \lk e^{\frac{x^2-\mu_f/T_0}{T/T_0}}+1\rk^{-1}. 
\eeq
\end{document}